\documentclass{iopart}

\usepackage{iopams}

\def\a{\alpha}
\def\b{\beta}

\def\de{\delta}   

\def\phi{\varphi}
\def\la{\lambda}

\def\s{\sigma}
\def\z{\zeta}

\def\vphi{\varphi}

\def\E{{\mathcal E}}

\def\R{{\bf R}}

\def\X{{\mathcal X}}
\def\Y{{\mathcal Y}}

\def\pa{\partial}
\def\pd{\partial}
\def\d{{\rm d}}       

\def\xd{{\dot x}}
\def\yd{{\dot y}}
\def\zd{{\dot z}}
\def\wdot{{\dot w}}

\def\grad{\nabla}     

\def\EOR{\hfill {$\odot$}}
\def\EOP{\hfill {$\triangle$}}

\def\fracor#1#2{({#1} / {#2} ) }

\def\({\left(}
\def\){\right)}
\def\=#1{\overline #1}

\def\~#1{\widetilde #1}
\def\wt#1{\widetilde #1}
\def\.#1{\dot #1}
\def\^#1{\widehat #1}

\def\beq{\begin{equation}}
\def\eeq{\end{equation}}
\def\eqref#1{(\ref{#1})}

\def\symmref{EMS1,CGbook,KrV,Olv1,Olv2,Ste}

\begin{document}

\title{Dynamical systems and $\s$-symmetries}

\author{G. Cicogna$^1$, G. Gaeta$^2$ and S. Walcher$^3$}

\address{{}$^1$ Dipartimento di Fisica, Universit\`a di Pisa, and INFN
sezione di Pisa, Largo B. Pontecorvo 3, 56127 Pisa (Italy); e-mail: {\tt
cicogna@df.unipi.it} \\
{}$^2$ Dipartimento di Matematica, Universit\`a degli Studi di Milano,
via C. Saldini 50, 20133 Milano (Italy); e-mail: {\tt
giuseppe.gaeta@unimi.it} \\
{}$^3$ Lehrstuhl A f\"ur Mathematik, RWTH Aachen,
52056 Aachen (Germany); e-mail: {\tt walcher@mathA.rtwh-aachen.de}}

\begin{abstract} A deformation of the standard prolongation operation,
defined on sets of vector fields in involution rather than on
single ones, was recently introduced and christened
``$\s$-prolongation'';
correspondingly one has ``$\s$-symmetries'' of differential equations.
These can be used
to reduce the equations under study, but the general reduction
procedure under $\s$-symmetries fails for equations of order one.
In this note we discuss how $\s$-symmetries can be used to reduce
dynamical systems, i.e. sets of first order ODEs in the form
$\dot{x}^a = f^a (x)$.

\end{abstract}

{\small \parskip=0pt \parindent=0pt

{\tt PACS numbers:} 02.30Hq

{\tt MSC numbers:} {34C14; 34A34, 34A05, 53A55}

{\tt Keywords:} Symmetry of differential equations; dynamical
systems; reduction; nonlinear systems }

\section*{Introduction}

In recent papers \cite{CGW,Sprol} we have introduced a generalized
prolongation operation, defined not on single vector fields but on
sets of vector fields (in involution), and depending on a smooth
matrix function $\s : J^1 M \to \mathrm{Mat} (n,\R)$.
This was called {\it $\s$-prolongation}, or
{\it joint $\la$-prolongation} to emphasize that it is an extension
of the $\la$-prolongation introduced a decade ago by Muriel and
Romero \cite{MuRom1} (see also \cite{MuRom2,MuRom3,MuRom3b,MuRom4,
MuRom5,MuRom6,MRO,SIG} and \cite{Gtwist,PuS}).
Correspondingly we introduced a notion of {\it $\s$-symmetry}, or {\it
joint $\la$-symmetry}, for systems of ODEs: Lie-point vector fields which
-- after being $\s$-prolonged up to suitable order -- leave a given set of
equations $\mathcal{E}$ invariant, are said to be {\it $\s$-symmetries}
for $\E$.

It was also shown \cite{Sprol} that $\s$-prolongations enjoy the
``invariants by differentiation property'', and hence they may be used,
essentially in the same way as standard (and $\la$ in the scalar
case \cite{MuRom3}) prolongations to reduce systems of ODEs of order $q
\ge 2$; see \cite{Sprol} for details.

These results left out the special -- but relevant -- case of {\it
dynamical systems} (DS in the following), i.e. of systems of first
order ODEs of the form
\beq\label{eq:DS0} d x^a / d t \ = \ F^a
(x,t) \ . \eeq
The purpose of this note is to fill this gap, i.e. to discuss how
$\s$-symmetries can be used in the study of DS and thus complete the discussion
of \cite{Sprol}.

Actually, by a standard procedure (based on adding a variable; see e.g.
\cite{CGW} for relations between symmetries and orbital
symmetries of the original and the modified system), one can
always reduce to consider autonomous DS. We will therefore deal
with equations of the form
\beq\label{eq:DS} d x^a / d t \ = \ f^a (x) \ \ \ \ (a=1,...,n) \ . \eeq
Similarly, we will later on restrict to consideration of
Lie-point time-independent vector fields,
$X_i = \vphi^a_i (x) (\pa / \pa x^a)$ (the Einstein summation convention
over repeated indices is used throughout this paper); we prefer however to
give in the next section a discussion for DS in their general, possibly
not autonomous form \eqref{eq:DS0}, and for general vector fields as well,
for convenience of the reader.

It should be mentioned that we already investigated applications of this
new type of symmetries to DS in a previous paper \cite{CGW}; in that
paper, however, we made no explicit use of prolongation for first order
systems, and rather emphasized the general framework and structural properties
(and correspondingly a more abstract mathematical approach).
Moreover,  in \cite{CGW} we were also interested in a strategy aiming at
writing at least some of the (first order) equations in higher order; this
would remove the degeneracies related to having a first
order system \cite{KrV,Olv1,Olv2,Ste} and hence will help in the search
for symmetries. Here we will {\it not } follow this line, but,
following an approach similar to the one taken in \cite{Sprol}
(which as said above was not able to deal with DS), we just
consider equations in their natural first order form; in other words, we
aim at a ``direct'' extension of the approach and results in \cite{Sprol}
in order to complete our study, with a view at
concrete reduction results procedures.

Also, in \cite{Sprol} we considered the geometrical aspects of our approach;
this will not be discussed here in order to avoid duplication of material:
the interested reader is referred to that paper (and similarly to
\cite{CGW}
for algebraic aspects as well). Again in \cite{Sprol} (see Sect.6 and
Appendix B in there), we discussed some
limitations of our approach from the practical point of view, due to the
difficulty of actually determining $\s$-symmetries of a given system.
In view of the relevance of this point for applications, we
report here some points of our discussion, but we also mention
how we could give constructive results in some cases.

Focusing on dynamical systems one can look at our approach as just a different way of 
looking at something which is already well-known, i.e. it is based on the very Frobenius reduction theory. 
It should be mentioned, however, that the same can be said for virtually any reduction procedure,
and that setting this in the language of Lie symmetries leads to a wider range of concrete applications.

Finally, we would also like to mention that, as already stressed in
\cite{Sprol} (see again there for further detail), our approach should be seen
as an extension and generalization of the approach by Pucci and Saccomandi
\cite{PuS} to the Muriel and Romero fruitful idea of $\la$-symmetries
\cite{MuRom1,MuRom2,MuRom3,MuRom3b}.

The {\it plan of the paper } is as follows. We will first recall the
definition and relevant properties of $\s$-prolongations, then prove a
theorem showing how they can be used to reduce
$\s$-symmetric DS and present some remarks about this reduction procedure.
We will also show that this is related to the existence of
$\sigma$-invariant constants of motion of the DS.
We then discuss the problem of determining $\sigma$-symmetries, which is in
general, as said before, beyond reach, apart from some ``favorable'' cases.
We also give a number of explicit examples 
in low dimensions, which illustrate the various situations, and finally in the end we draw some conclusions.

All the objects (manifolds, functions, vector fields) considered
in this paper will be assumed to be smooth.


\section{Definition and properties of $\s$-prolongations, and
$\s$-symmetries.}
\label{sec:definition}

We denote as usual by $M = \R \times U$ the extended phase
manifold for the DS \eqref{eq:DS}, where $t \in \R$ and $x \in U$;
here $U$ is a smooth manifold of dimension $n$. As our
considerations will be local, we can assume $U \simeq \R^n$ with
no loss of generality; on the other hand, in this section we will
consider general vector fields on $M$.

Consider a set $\X = \{ X_1 , ... , X_s \}$ of smooth vector
fields on $M$; assume they are in involution and their rank
(assumed to be constant for the sake of simplicity)
$ r \le s$ satisfies $r < n$. The involution
assumption means there are smooth functions $\mu_{ij}^k : M \to
\R$ such that
\beq [X_i,X_j] \ = \ \mu_{ij}^k \, X_k \ . \eeq

\subsection{Standard prolongations and symmetries}

As well known \cite{\symmref}, one associates to $M$ the Jet
manifold $J^1 M$; the vector fields $X_i$ are naturally lifted (or
{\it prolonged}) to vector fields $Y_i$ in $J^1 M$.

If the $X_i$ are written in local coordinates $(t;x)$ as
\[ X_i \
= \ \xi_i (x,t) \, \frac{\pa}{\pa t} \ + \ \vphi^a_i (x,t) \,
\frac{\pa}{\pa x^a} \ , \]
then the (first prolongations) $Y_i$ are written in the
natural local coordinates $(t;x;\xd)$ on $J^1 M$ as
\[ Y_i \ = \ \xi_i (x,t) \, \frac{\pa}{\pa t} \ + \
\vphi^a_i (x,t) \, \frac{\pa}{\pa x^a} \ + \ \psi^a_i \,
\frac{\pa}{\pa \xd^a} \ . \]
When $Y_i$ is the standard
prolongation of $X_i$, the coefficients $\psi^a_i$ are given by
the (standard) prolongation formula
\[ \psi^a_i \ = \ D_t \vphi^a_i \ - \ \xd^a \, D_t \xi_i \ . \]

The vector field $X$ is said to be a {\it symmetry} for the
equation $\E$ defined on $J^1 M$ if
$[Y(\E)]_{S(\E)} = 0$, where
$S(\E)$ is the {\it solution manifold} for $\E$, i.e.
the set of
points of $J^1 M$ on which $\E$ is satisfied. If the relation is
satisfied without restricting to $S(\E)$, i.e. if $Y(\E) = 0$
on all of $J^1 M$, then $X$ is said to be a {\it strong symmetry}. It is
well
known that if $\E$ admits $X$ as symmetry, then there is an equivalent
equation $\wt{\E}$ (i.e. an equation $\wt{\E}$ with the same
solutions as $\E$) which admits $X$ as a strong symmetry
(see e.g. \cite{CDW}); we stress that this result is obtained by working
purely in $J^1 M$, hence it can be rephrased by saying that if $\E$
is invariant under $Y$, then there is $\wt{\E}$ which is strongly
invariant under $Y$ (independently of the fact that $Y$ is the
prolongation of~$X$).

\subsection{$\s$-prolongations and $\s$-symmetries}

The $\s$-prolongation represents a modification of the standard
prolongation operation. The relation between the $\psi$ and the
$\vphi$ coefficients is in this case given by
\beq\label{eq:sprol}
\psi^a_i \ = \ \( D_t \vphi^a_i \ - \ \xd^a \, D_t \xi_i \) \ + \
\s_{ij} \, \( \vphi^a_j \ - \ \xd^a \, \xi_j \) \ , \eeq
with
$\s_{ij} : J^1 M \to \R$ smooth functions; this modified
prolongation operation is thus characterized by the smooth $s\times s$
matrix function $\s$.

Note that $\s$-prolongation is defined on sets of vector fields,
not on individual ones, as $\psi^a_i$ depends on the coefficients
$\vphi^a_j$ for the other vector fields. When the set consists of
a single vector field, $\s$-prolongation reduces to the
$\la$-prolongations of Muriel and Romero
\cite{MuRom1,MuRom2,MuRom3,MuRom3b} (see also \cite{Gtwist}).

It should also be noted that for a generic choice of $\s$, the
$Y_i$ will {\it not } be in involution even if the $X_i$ are.
Sufficient conditions for the involution properties to be
preserved under $\s$-prolongation (in the form of an equation to
be satisfied by $\s$) are discussed in \cite{Sprol}. We will here
{\it assume } that the $Y_i$ are in involution.

Note that it may happen that the $Y_i$ are not in involution, but
can be completed to an involution system by adding a finite number
of vector fields; if the rank $\rho > r$ of the completed system
satisfies $\rho < n$, the reduction to be discussed below is still
possible with suitable modifications (see Section
\ref{sec:completion}).

If the $\s$-prolongations $Y_i$ of the $X_i$ leave the system of
equations $\E$ invariant, i.e. if
$[Y_i (\E)]_{S(\E)} = 0$,
we say that the involution system $\X = \{ X_i \}$ is a {\it
$\s$-symmetry} for $\E$.
If $Y_i (\E) = 0$, i.e. if the invariance condition holds
without restriction to $S(\E)$, we say that $\X$ is a
{\it strong $\s$-symmetry } for $\E$.
The same result mentioned above holding for standard and strong
symmetries is immediately seen to hold for $\s$-symmetries and
strong $\s$-symmetries: if $\E$ admits $\X$ as $\s$-symmetry, then
there is an equivalent system $\wt{\E}$ admitting $\X$ as a strong
$\s$-symmetry. In the following we will thus deal with
strong symmetries.

\subsection{The ``invariants by differentiation'' property}

It is well known that for standard prolongations one can generate
differential invariants of all orders starting from those of lower
order. This is based on the so-called ``invariants by
differentiation property'' (IBDP).

The same property extends to $\la$-prolongations
\cite{Gtwist,MuRom1,MuRom2,MuRom3,MuRom3b} (and can be used to
characterize them
\cite{PuS}), and also to $\s$-prolongations. For these we have
\cite{Sprol} that, with $D_t$ the total derivative with respect to time
$t$, if the $\Y$ are $\s$-prolonged, they satisfy
\beq [ Y_i ,
D_t ] \ = \ \s_{ij} \, Y_j \ - \ \( D_t \xi_i \, + \,
\s_{ij} \, \xi_j \) \, D_t  \ . \eeq

\medskip\noindent
{\bf Proposition 1} \cite{Sprol}. {\it Let $\Y$ be a set of
$\s$-prolonged vector fields, and let $\zeta_1,\zeta_2$ be common
independent differential invariants of order $k$ for all of them. Then
\beq\label{eq:Theta} \Theta \ := \ (D_t \zeta_1)/(D_t \zeta_2) \eeq
is a common differential invariant of order $k+1$ for all of them.}
\bigskip

Note that in particular, if the $X_i$ are ``vertical'' vector fields, i.e.
if $\xi_i=0$ (see next section and in particular Remark 1 for a
discussion on this point),  then necessarily $t$ is an
invariant for the $X_i$ (and hence also for the $Y_i$), and
Proposition 1 makes that given an invariant $\zeta = \zeta (x)$ of
order zero, we obtain immediately (just choose $\zeta_2 = t$ in
\eqref{eq:Theta}) a first order differential invariant in the form
\beq \zeta^{(1)} \ := \ D_t \, \zeta \ . \eeq

\section{Reduction of DS and $\s$-symmetries}
\label{sec:reduction}

The reduction procedure discussed in \cite{Sprol} for systems of
differential equations of order $q>1$ does not apply to the
special case of dynamical systems. The obstacle lies in that the
most general differential invariants of order $q$ can be obtained,
for $q>1$, from those of lower order (thanks to the IBDP) and in
particular from those of order zero and one (the need to include
differential invariants of order one is intuitively clear, as at
order zero there is no trace of the $\s$ matrix in the vector
fields). But for $q=1$ this approach fails.

We should therefore consider the special case $q=1$, and in
particular that of dynamical systems (thus, equations solved with
respect to first derivatives of dependent variables), with a
different approach.

As anticipated in the Introduction, we will
consider autonomous DS \eqref{eq:DS} and correspondingly restrict
to consideration of Lie-point time-independent vector fields, $X_i
= \vphi^a_i (x)(\pa / \pa x^a)$.

\medskip\noindent
{\bf Remark 1.} Note in this respect that, as well known, when dealing
with a general vector field $X = \xi (\pa / \pa t) + \phi^a (\pa /
\pa x^a)$, one can equivalently pass to consider its {\it evolutionary
representative} $X_{ev} = (\phi^a - \xi \xd^a) (\pa / \pa x^a)$
\cite{Olv1,Olv2,Ste}. Upon restriction to solutions to \eqref{eq:DS}, this
reads as $X_{ev} = (\phi^a - \xi f^a) (\pa / \pa x^a)$; the latter is a
well defined vector field in $M$, and of the form we consider.  We can also
proceed in a slightly different way: on the set $S = S(\E)$ of
solutions to \eqref{eq:DS}, one has
$[\pa / \pa t]_S\! \equiv\!- f^a (\pa / \pa x^a)$, and hence $X$ is again rewritten as
$[X]_S \equiv (\phi^a - \xi f^a ) \pa / \pa x^a$. \EOR

\medskip
In the assumptions $\xi_i=0$ and $\phi^a_i$ independent of time,  the
$\s$-determining equations, i.e. the condition for the DS \eqref{eq:DS} to
be invariant
under the $\s$-prolongations $Y_i$, where
\[Y_i\ =\ \varphi_i^a\frac{\pa}{\pa
x^a}+(D_t\vphi_i^a+\s_{ij}\vphi_j^a)\frac{\pa}{\pa\dot x^a}\]
takes the form of   the Lie bracket condition
\beq\label{eq:Ssymcom} [ X_i , X_0 ] \ = \ \s_{ij} \ X_j  \quad\quad
   (i,j=1,\ldots,s)
\eeq
where $X_0  =  f^a (x) \pa/\pa x^a$ is the vector field describing the
dynamics (see also subsect. 5.1 for some detail).

\medskip\noindent
{\bf Remark 2.} One can look at this construction in a different
way, which is established in the literature and goes as follows.
Since the vector fields $\{ X_1 , ... , X_s \}$ are in involution,
they define a foliation (regular due to our hypotheses). Condition
\eqref{eq:Ssymcom}  implies that the vector field $X_0$ is
projectable with respect to this foliation (see e.g.
\cite{MM}, pp. 29 ff.). Conversely, given a foliation and a
projectable vector field $X_0$, consider a set of vector fields
$X_i$ whose common invariants define -- by means of their level
sets -- the leaves of the foliation; then the 
$\{ X_0 ; X_1 , ... , X_s \}$ will satisfy a set of relations as in
\eqref{eq:Ssymcom}.
It may be noted that such constructions are well known in symmetry
reduction, see e.g. \cite{Her}. For the special case of
Hamiltonian systems the leaves are the level sets of the momentum
map, see \cite{OrR}. In this sense, the results contained in
this paper, albeit obtained in a different way, can be seen as a
recasting of this classical approach (basically going back to Frobenius)
in the language of Lie symmetries. \EOR

\medskip\noindent
{\bf Remark 3.}
We note that our general concept of reducibility covers the
slightly broader setting of {\it orbital} reduction: in this case in
identity \eqref{eq:Ssymcom} the summation on the r.h.s. ranges
from 0 to $s$, or, in explicit form,
\beq\label{eq:Sorb} [ X_i , X_0 ] \ = \ \s_{i0}X_0+\s_{ij} \ X_j  \quad\quad
   (i,j=1,\ldots,s)
\eeq
with some $\s_{i0}\not=0$. In this case, the zero-order common invariants
$\zeta_i\ (i=1,\ldots, n-r)$ satisfy an {\it orbitally reducible}
system, i.e. a system ``reducible up to a common scalar factor'':
\[ \dot\zeta_i\ =\ \rho(x)\,f_i(\zeta)\ .\]
See \cite{CGW} for a proof and more detail. \EOR

\medskip\noindent
{\bf Theorem 1.} {\it Let $\mathcal{X}$ be a set of vector fields on the
$n$-dimensional manifold $M$; assume $\mathcal{X}$ is in involution and of
rank $r < n$. Let $\mathcal{Y}$ be the $\sigma$-prolongation of
$\mathcal{X}$, and let $\mathcal{Y}$ be in involution and also of rank
$r$. If the $n$-dimensional dynamical system (2) is invariant under the
set $\mathcal{Y}$, then the system can be locally reduced, passing to
suitable symmetry-adapted coordinates, to a dynamical system of dimension
$n-r$ and a system of $r$ ``reconstruction equations'' depending on the
solution to the reduced system.}

\medskip\noindent
{\bf Proof.} Let us first consider the situation at order zero,
i.e. for the action of the vector fields $X_i$ in $M$; the latter
is of dimension $n+1$ and hence we will have, apart from the
trivial invariant $\tau = t$, other $n-r$ independent common invariants
$\zeta_i
(x)$ as guaranteed by Frobenius theorem. We can consider a change
of coordinates in $M$ and the set of coordinates $z_i = \zeta_i
(x)$ ($i=1,...,n-r$) will be complemented by some $r$ coordinates $y_j
= \eta_j (x)$ ($j=1,...,r$). The vector fields $X_i$ will be
written, in the new coordinates, as
\[ X_i \ = \ \Phi^j_i \ (\pa / \pa y^j) \ . \]
Let us now consider the situation in $J^1 M$ (of
dimension $2 n + 1$) with the action of the first
$\s$-prolongations $Y_i$. Now we have $(2n +1 - r)$ invariants,
with $n$ of them being genuinely of first order. By Proposition 1,
we know that $\zeta_i^{(1)} := D_t \zeta_i$ is a first order
invariant for the $Y_i$, and this allows to identify $(n-r)$ such
invariants. There will be other $r$ first order invariants,
$\beta_j$, which are not obtained in this way; these will involve
the $D_t \eta$ and, in general, can also depend on the $\zeta ,
D_t \zeta$ (note however that any dependence on the $\dot{z}$ can
be eliminated e.g. by Gram-Schmidt--type procedure). The $Y_i$
will be written, in the new coordinates, as
\[ Y_i \ = \ \Phi^j_i
(\pa / \pa y^j) \ + \ \Psi^a_i (\pa / \pa \dot{y}^j) \ . \]

We can then pass to consider, in $J^1 M$, the system of
coordinates $(z,y ; \dot{z} , \dot{y})$. The equations of motion
for the $z$ and $y$ will be obtained by rewriting \eqref{eq:DS} in
the new coordinates. In full generality, these would read
\[
\dot{z}_i \ = \ f_i (z,w) \ ; \ \ \dot{y}_j \ = \ g_j (z,y) \ . \]
However, the $Y$ are symmetries of the equations, and the
$\dot{z}_i$ are invariant under the $Y$; it follows that the $f_i$
are also invariant, i.e. we should have $f_i = f_i (z)$. This
shows that we get a reduced dynamical system, of dimension $n-r$,
for the invariant variables
\beq\label{eq:DSz} d z_i / d t \ = \
f_i (z ) \ \ \ \ (i = 1 , ... , n-r) \ . \eeq

As for the other $r$ equations, $\dot{y}_j = g_j (z,y)$ they must
also be invariant under $Y$ (at least when restricting to the
solution of the equations \eqref{eq:DSz}), and hence can be
written in terms of the first order invariants; in full
generality, they can be given the form $\beta_i = b_i$ where $b_i$
depends only on the $z$ and $\dot{z}$ invariants; restricting to
the solutions of \eqref{eq:DSz} means that we can write $f_i (z)$
for $\dot{z}_i$, and hence these equations are written in the form
\beq\label{eq:DSbeta} \beta_j \ = \ B_j (z) \ ; \eeq
the exact
form of the functions $B_j$ depends of course on the initial
equations \eqref{eq:DS} and on the arbitrary choices made in the
definition of $\beta_j$ (they involve coefficients being function
of $z$ as well).

The set \eqref{eq:DSz} represents the reduced system, and the
\eqref{eq:DSbeta} are the reconstruction equations; they can be
thought as a set of (generally coupled) not autonomous equations
for the $y(t)$, depending on the solution $z_i (t)$ of the reduced
equations. \EOP

\medskip\noindent
{\bf Remark 4.} The above Theorem is stated and proved
in a quite different setting and perspective in \cite{CGW}, where
reduction procedures of dynamical systems, involving also
orbital symmetries, with extensions to ODE's of higher order, are discussed
focusing on more algebraic aspects,  generalizing in several ways the
approach and the results of \cite{HaWa,Wal99}.
 \EOR

\medskip\noindent
{\bf Remark 5.} Here the ``reconstruction equations'' will in
general be (a system of coupled) differential equations in the
original variables, as they involve differential invariants. Thus,
at difference with the reconstruction procedure met when using
standard symmetries (which requires only quadratures), solving the
reconstruction equations is in general a nontrivial task, and can
turn out to be impossible. We are not aware of any reasonably
general and/or natural condition ensuring that the reconstruction
equations can be solved. In other words, the reduction based on
$\s$-symmetries is a possible strategy to attempt in dealing with
nonlinear systems, but with no guarantee of success even when one
is able to determine (some of) the system's $\s$-symmetries. On
the other hand, there are cases (as will be shown in the Examples
below) where one is able to solve the reconstruction equations.
\EOR

\medskip\noindent
{\bf Remark 6.} We stress that the situation described above
for $\s$-symmetries was already present when dealing with
so-called $\Lambda$-symmetries \cite{Cds} (also called
$\rho$-symmetries \cite{Cds,Gtwist}); these are $\mu$-symmetries
\cite{CGMor} for the specific setting
of dynamical systems \cite{Gtwist} and turn out to be a direct
generalization of $\la$-symmetries \cite{MuRom1,MuRom2,MuRom3,MuRom3b}.
\EOR

\medskip\noindent
{\bf Remark 7.} Our discussion shows that, as it often happens with
symmetry considerations, $\s$-symmetry properties are specially
transparent when adopting symmetry adapted coordinates. Some of these
coordinates should correspond with invariants, while our discussion left
the choice of other coordinates free. A specially convenient  choice would
be that of coordinates which rectify $r$  vector fields $\{ X_1 ,
... , X_r \}\ (r\le s)$ in the set $\X$; then the $Y_i$ take
the particularly simple and significant
form
$Y_i = ( \pa / \pa y^i) + \s^{ij} (\pa / \pa
\yd^j) \  (i,j = 1,...,r) $,
which extends a similar result  for $\la$-symmetries \cite{MuRom3b}.
One may again see this in the light of Frobenius theorem and related
reduction approaches, see Remark 2.
\EOR

\medskip\noindent
{\bf Remark 8.} As mentioned above, the reduction obtained through
the procedure we are considering is the restriction of the initial
system to the space of invariants (under the $\s$-symmetries). In
this sense, the present approach can be seen as the extension to
fully general DS of an approach (based on Michel theory
\cite{Mic1,Mic2,Mic3}) developed some time ago for systems in
normal form \cite{Gnf,GWnf1,GWnf2} (see also \cite{GWnf3}).
\EOR

\section{Completion of involution system}
\label{sec:completion}

The formulation of Theorem 1 requiring that $\X$ and $\Y$ have the same
rank, does not consider a case which can
occur in applications: there is in fact the possibility that the
involution properties of the vector fields $X_i$ are different
from those of the $Y_i$.

This may seem a contradictory statement: in fact, it is clear that
\beq\label{eq:invY} [Y_i , Y_j ] \ = \ \mu_{ij}^k \, Y_k \eeq
(note here the $\mu_{ij}^k$ are in general functions on $J^1 M$
with values in $\R$) requires also
\beq\label{eq:invX} [X_i , X_j ] \ = \ \mu_{ij}^k \, X_k \ . \eeq

On the other hand, a little thinking (or some explicit examples,
see below) show that it is well possible that starting from an
involution system \eqref{eq:invX} and applying the
$\s$-prolongation procedure one obtains a set of vector fields
$Y_i$ which do not satisfy \eqref{eq:invY}; actually this will be
the generic case for a randomly chosen $\s$. When \eqref{eq:invY}
is not satisfied, the $\{ Y_i \}$ could either not close to a
nontrivial involution system (that is, more precisely, only close
once vector fields along all directions in $J^1 M$ are added to
the system), or close to an involution system $\overline\Y$ after
adding a certain number of auxiliary vector field, still however
providing a system of rank $r < n$. In this case Theorem 1 (which
only makes reference to the prolonged vector fields, i.e. to the
vector fields in $J^1 M$) still applies.

Some further considerations are in order regarding this case. As
remarked above, the involution properties  \eqref{eq:invY} among
the $Y_i$ imply the same involution properties \eqref{eq:invX} are
satisfied by the $X_i$. If we start from the latter, it is clear
that the completion procedure (i.e. adding to the set any vector
field appearing in the commutators $[Y_i,Y_j]$ and so on) will
produce only vector fields which have zero projection in $M$ (i.e.
they are vertical for the fibration $J^1 M \to M$). In other
words, the new vector fields will be written in coordinates as
\beq\label{eq:auxY} Y_m \ = \ \psi^i_m \ \fracor{\pa}{\pa \xd^i} \
. \eeq

Vector fields of this form can be seen as $\^\s$-prolongations
(here $\^\s$ is an extension of the original matrix $\s$, see
below) of the trivial vector fields $X_m = 0$ in $M$. Needless to
say, such vector fields commute with all the $X_i$, and hence the
auxiliary vector fields $Y_m$ will also commute with all the
$Y_i$, and will be an abelian subalgebra in the center of the Lie
algebra $\overline\Y$.

As for the matrix $\^\s$, this will be written in block form as
\[ \^\s \ = \ \pmatrix{\s & A \cr \rho & B \cr} \ , \]
where $\rho$ embodies the relation between the $\psi_m$ and the
$\phi_i$, while $A$ and $B$ are arbitrary (they act on the null
$\phi_m$ vectors) and can be set to zero; if we require to invert
$\^\s$ (as happens when discussing the gauge meaning of
$\s$-prolongations, see \cite{Sprol}) we can equally well set
$A=0$, $B=I$.

Finally, note that the presence in $\overline\Y$ of vector fields
of the form \eqref{eq:auxY} can forbid the presence of some
$\xd_i$ in the differential invariants (this is specially clear if
$\psi_m^i$ are constant); correspondingly, in this case some of
the variables in $M_0$ do actually play the role of parameters. In
other words, the invariant DS can not have full dimension in
$M_0$, and we will be dealing with DS of dimension $n_0 < n$. The
number of parameters $\de = n - n_0$ corresponds, generically, to
the number of auxiliary vector fields \eqref{eq:auxY} that must be
introduced to complete the involutory system, and more precisely
to $\mathtt{rank} (\overline\Y ) - \mathtt{rank} (\Y)$.

In this case it will be more convenient to distinguish between
real dynamical variables and parameters; we will thus consider $M
= M_0 \times P$, where $M_0$ represents the phase manifold and $P$
the parameter space. The vector fields $X_i$ (and hence $Y_i$)
will be allowed to depend on parameters and act on them.

It turns out that also in the case one needs to complete
$\s$-prolongations $\Y$ in order to have an involution system
$\overline\Y$, if the rank of $\overline\Y$ is sufficiently small,
we can still perform the reduction. We will now give a more
precise description of the reduction procedure in this framework.

\medskip\noindent
{\bf Theorem 2.} {\it Consider an $n_0$-dimensional autonomous DS
\eqref{eq:DS} in $M_0 = \R^n$, with $n_0 \le n$. Assume it admits
the set $\X = \{ X_1 , ... X_m \}$ of vector fields in $M_0$, of
constant rank $r_0 < n$, as $\s$-symmetries. Assume moreover that the
completion $\overline\Y$ of the set $\Y$ corresponding to the
$\s$-prolongation of $\X$ has rank $r = r_0 + \de$ (with $0 \le r < 2 n$)
in $J^1
M$. Then \eqref{eq:DS} can be reduced, passing to suitable
symmetry-adapted coordinates, to an autonomous DS of dimension
$\kappa_0 = n-r_0$ and to a set of $\theta = r_0 - \de$
``reconstruction equations'' depending on the solutions to the
reduced system.}

\medskip\noindent
{\bf Proof.} As mentioned in the statement, we will consider only
autonomous DS and autonomous vector fields; as mentioned above one
can always reduce to this setting by the standard procedure of
autonomization -- i.e. adding a new variable $x_0$ corresponding
to $t$ and satisfying $\xd_0 = 1$.

Let the commutation properties between the $X_i$ be described by
\eqref{eq:invX}, and let $\mathtt{rank} (\X) = r_0 < n$. Thus
there are $\kappa_0 = n - r_0 $ nontrivial invariants $\z_i$ of
order zero in the $(n+1)$ dimensional phase manifold $M$ (plus one
trivial invariant, corresponding to $t$). By the IBDP, these
generate $\kappa_0$ differential invariants $\z_i^{(1)}$ of order
one, which will be referred to as derived invariants.

On the other hand, the vector fields $Y_i$ act in $J^1 M$, which
has dimension $(2n + 1)$. By assumption, the set $\overline\Y$ has
rank $r \ge r_0$; we set $r = r_0 + \de$. Hence $\overline\Y$
admits $\kappa_1 = 2n + 1 - r$ invariants in $J^1 M$; Of these,
$\kappa_0 + 1$ are invariants of order zero, and $\kappa_0$ are
derived invariants. Thus there are $\theta$ additional invariants
$\beta_i$, with
\[ \theta \ = \ (2n + 1 - r) \ - \ (2 \kappa_0 + 1) \ = \ r_0 \ - \ \de \
. \]
As the $\Y$-symmetric system can be written in terms of
invariants, we can write it as a DS of dimension $\kappa_1$; there
are $\kappa_0 = n - r_0$ equations corresponding to derived
invariants, and hence of the form
\[ d \zeta_i / d t \ = \ \Phi_i
(\zeta_1 , ... , \zeta_{\kappa_0} ) \ , \]
and $\theta = r_0 -
\de$ reconstruction equations, corresponding to
\[ \beta_i \ = \ h_i(\z_1 , ... , \z_{\kappa_0} ) \ , \]
as stated. This completes the proof. \EOP


\section{Constants of motion and $\s$-symmetries}
\label{sec:CoM}

In the study of dynamical system, a special attention is given to the
search for {\it constants of motion}, i.e. of functions $I : M_0 \to \R$
which are constants under the flow of the DS; if the latter is written in
the form \eqref{eq:DS}, one is looking for functions such that
\[ D_t  I\ = \ \( \frac{\pa I}{\pa x^a} \) \ f^a (x) \ = \ 0 \
. \]

It is well known that strict  relations exist between standard symmetries
of a
DS and its constants of motion; a relation also exists between
$\s$-symmetries of a DS and its constants of motion.

\medskip\noindent
{\bf Theorem 3.} {\it Let the DS \eqref{eq:DS} admit a set $\X$ of
$\s$-symmetries of rank $r$; then it admits  $n - r -1$
independent constants of motion which are simultaneously invariant under
the $\s$-symmetry vector fields $X_i$.}

\medskip\noindent
{\bf Proof.}
The $\sigma$-symmetry condition in the form \eqref{eq:Ssymcom}
for  $\X=\{ X_1 , ... , X_s \}$, $X_i = \phi^a_i (\pa / \pa x^a)$, gives that
the enlarged set $\^\X = X_0 \cup \X = \{ X_0 ; X_1 , ... , X_s \} $ is a set
of $\^s = s+1$ vector fields in involution in the manifold $M_0$ of
dimension $n$, and its rank $\^r$ satisfies $r \le \^r \le r+1$
(generically, $\^r = r+1$). This set of vector fields will span an
$\^r$-dimensional distribution and hence will have at least
$n - \^r$ independent common invariants. But the invariance under $X_0$
expresses just the property of being a constant of motion of the
corresponding DS.
\EOP

\medskip\noindent
{\bf Remark 9.} Let us consider separately, for completeness, the case in
which the DS \eqref{eq:DS}  is actually the autonomized form of a not
autonomous initial DS \eqref{eq:DS0}. This means that \eqref{eq:DS}
contains the additional equation $\.x_0=1$ for the new variable $x_0$,
whereas the original DS \eqref{eq:DS0} involves $n-1=\^n$ dependent
variables $x_j(t)$ (and $n-1=\^n$ equations of course). Then, Theorem 1
ensures that there are $\^n+1-r$ reduced equations, or just $\^n-r$
reduced equations for the $\^n$ variables $x_j(t)$. Similarly, according
to Theorem 3, there are $\^n-r$ independent symmetry-invariant constants
of motion, or  (excluding the trivial one $x_0-t$) just $\^n-r-1$
constants of motion, which may depend on time. If instead the initial DS
\eqref{eq:DS0} is autonomous and the coefficient functions $\phi_i$ of
the vector fields $X_i$ are independent of time,  then also the constants
of motion provided by the above Theorem turn out to be independent of
time. \EOR

\medskip\noindent
{\bf Remark 10.} To illustrate an aspect of Theorem 3, consider for
comparison the special case
of a {\it Hamiltonian} DS, i.e. a DS of the form $\.x=J\grad_x H$, where
$J$
is the standard symplectic matrix and $H$ a given Hamiltonian (independent
of time, for simplicity). Assume that a vector field $X$ admits   a {\it
generating function} $G$, i.e. that $X=J\grad_x G$, and that $X$ is a
$\la$-symmetry for the DS. Then $G$ is in general {\it not} a constant of
motion for the DS \cite{Cds} (although it is trivially invariant under
$X$, and -- as well known -- a constant of motion if $\la=0$). Theorem 3
ensures that there are other constants of motion which are invariant under
$X$: in the presence of $m$ degrees of freedom (i.e. $n=2m$) and of just
one vector field $X$, the expected number of these constants of motion is
then $2(m-1)$. \EOR

\section{Determination of $\s$-symmetries}
\label{sec:det}

In this Section we would like to briefly discuss some points related to
the determination of $\s$-symmetries for a given DS (see also
\cite{CGW,Sprol} for a discussion on the general case). We also give
constructive results about $\s$-symmetries of a class of DS, see
next subsection.

\subsection{General considerations}
\label{sec:det_gen}

The determination of all the $\s$-symmetries for a given DS is in general
beyond reach; this should not be surprising as also the determination of
standard symmetries may be far from easy. In the same way as for standard
symmetries, however,
even a partial knowledge of the symmetry structure can be useful in that
it allows for a reduction of the system under study; thus one can look for
special $\s$-symmetries, e.g. by an educated guess in view of the features
of the system under study.

In general, a given set of vector fields $X_i = \xi_i (x,t) (\pa / \pa t )
+ \phi^a_i (x,t)
(\pa / \pa x^a)$, with $\s$-prolongation $Y_i = X_i + \psi^a_i (\pa / \pa
\xd^a)$, is a $\s$-symmetry of the DS \eqref{eq:DS} if
the $\s$-determining equations
\beq\label{eq:sdeteqgen} \psi^a_i \ - \ \phi^b_i \, (\pa f^a / \pa x^b) \
= \ 0 \eeq
are satisfied on the solutions to \eqref{eq:DS} itself. These
are a system of $n\cdot s$ equations for the unknown vector
fields $X_i$ (i.e. for the coefficient functions $\xi_i (t,x)$ and
$\phi_i (t,x)$ appearing in them) {\it and} for the $s^2$ components of
the matrix $\s = \s (t,x,\xd)$.

If we restrict to vector fields of the form $ X_i \ = \ \phi^a_i
(x)(\pa/\pa x^a)$,
the $\s$-determining equations take the form \eqref{eq:Ssymcom}, which
can be now more conveniently rewritten in component form
\beq\label{eq:sdeteq} \overline{\s}_{ij} \ \phi^a_j \ = \ \phi^b_i \, (\pa
f^a / \pa x^b) \ - \ f^b \, (\pa \phi^a_i / \pa x^b)  \eeq
where we have written
$\overline{\s}$ to emphasize that $\s$ is the restriction of
$\s (t,x,\xd)$ to the
solution manifold, i.e. $\overline{\s} (t,x) = \s (t,x,f(x))$.
These are again a system of equations for the unknown vector
fields $X_i$, i.e. for the $\phi^a_i (x)$, {\it and} for the $s^2$
components of
the $s \times s$ matrix function $\overline{\s}$. Thus we have strongly
under-determined systems, and one could consider further {\it ansatzes} in
searching for solutions. (Note that even for $s = 1$ (so for
$\la$-symmetries) we have a system of $n$ equations for $n+1$ unknown
functions, i.e. the $\phi^a$ and the (now, scalar) $\s$.)

There seems to be no algorithmic way to solve the systems
\eqref{eq:sdeteqgen} or \eqref{eq:sdeteq}; in practice one can look for
solutions to it only by making simplifying assumptions on the functional
form of the unknown functions $\xi_i,
\phi^a_i$ and $\s_{ij}$, or when the DS under study has a specially
favorable structure, see below.

\medskip\noindent
{\bf Remark 11.}
In this sense the new tool in symmetry analysis of DS we propose here is
quite different from standard Lie-point symmetry analysis, in that it is
{\it not }  algorithmic; it should be stressed that this is not a
degeneration of the DS case, and the same holds for higher order equations
\cite{Sprol} (so from this point of view nothing is gained by transforming
the DS into a higher order form, if this is possible). Once again this
feature is shared with other (related) extensions of Lie-point symmetries,
such as $\lambda$ and $\mu$-symmetries.
\EOR

\medskip\noindent
{\bf Remark 12.}
On the other hand, if we fix $X_i$ and $\s_{ij}$ (assuming the
$\s$-prolongation of the set $\X$ gives a set $\Y$ in involution; see the
discussion in previous sections), determining the most general DS
admitting the $\X$ as a $\s$-symmetry is in principle feasible, albeit
sometimes not easy in practice; for its solution one should take into
account the commutation properties of the system $\X$. The possibility of
determining the most general DS admitting a given involution set $\X$ as
$\s$ symmetry (for a given $\s$) is shown in some of the examples below.
\EOR

\medskip\noindent
{\bf Remark 13.} With reference to Remark 2 above, note that the
discussion of this section does also provide constructive tools to
determine foliations with respect to which a given vector field is
projectable.\EOR


\subsection{Special structure of the dynamical system}
\label{sec:det_gp}

We want now to discuss how a favorable structure of the DS under study can
lead to a tractable problem for the determination of $\s$-symmetries of a
given system.

We will consider DS in $\R^n$ of the form (cf. \cite{CGW})
\beq\label{eq:gpds1} \xd^a \ = \ F^a (x) \ = \ f^a (x) \ + \ \sum_{k=1}^s
\, \a_k (x) \ \phi^a_k (x) \ , \eeq
where $\alpha_k(x)$ are arbitrary functions,
with the property that the ``simplified'' system
\beq\label{eq:gpds2} \xd^a \ = \ f^a (x) \eeq
admits the system $\X = \{ X_i , ... , X_s \}$ of vector fields
\beq\label{eq:gpvf1} X_i \ = \ \phi^a_i \ (\pa / \pa x^a ) \eeq
as standard symmetries, and that the $\{ X_1 , ... , X_s \}$ thus defined
are in involution,
\beq\label{eq:gpvf2} [ X_i \, , \, X_j ] \ = \ \b_{ij}^k \ X_k \ . \eeq

\medskip\noindent
{\bf Theorem 4.} {\it Let the DS \eqref{eq:gpds2} admit the set $\X$ of
vector fields \eqref{eq:gpvf1}, satisfying \eqref{eq:gpvf2}, as standard
symmetries. Assume in particular, for simplicity, that $\b_{ij}^k$ are constant, and
consider the set $\Y$ of vector fields obtained as $\sigma$-prolongation
of the set $\X$   with $\sigma$ given by
\beq\label{eq:gpsigma} \s_{ij} \ = \ \a_k \ \b_{ik}^j \ + \ X_i (\a_j ) \
. \eeq
Then: i) the  $Y_i$ are in involution
and  actually satisfy the same involution properties as the $X_i$; ii)
any DS of the form \eqref{eq:gpds1} admits ${{\cal X}}$, with $\sigma$
given by \eqref{eq:gpsigma}, as a $\sigma$-symmetry
and therefore can be reduced according to Theorem 1.}

\medskip\noindent
{\bf Proof.} We have shown in \cite{Sprol} (see Corollary 2 there)
that a sufficient (but by no means necessary) condition for the
set of $\s$-prolonged vector fields $\Y$ to be in involution, and actually
satisfy the same involution relations as the original set $\X$, is that
$\s$ satisfies a certain equation. In the present case, i.e. if $\s$ is of
the form \eqref{eq:gpsigma},
and the $\b_{ij}^k$ are constant, this equation reduces -- in the present
notation -- to
\[ X_i (\s_{jk} ) - X_j (\s_{ik} ) \, + \, \( \s_{im} \b_{mj}^k - \s_{jm}
\b_{mi}^k  -  \b_{ij}^m  \s_{mk} \) \ = \ 0 \ . \]
This is always satisfied for $\s$ as in \eqref{eq:gpsigma}. In fact,   now
$ X_m (\b_{ij}^k ) = 0$ for all choices of $i,j,k,m$, so that using
\eqref{eq:gpsigma} and recalling \eqref{eq:gpvf2},
the above equation reads
$$\begin{array}{l} \( \b_{jm}^k X_i (\a_m ) - \b_{im}^k X_j (\a_m) +
\b_{ij}^m X_m (\a_k ) \) \ + \ \a_\ell \left[ \b_{i \ell}^m \b_{m j}^k -
\b_{j
\ell}^m \b_{m i}^m - \b_{i j}^m \b_{m \ell}^k \right] \\
\ + \ \( X_i (\a_m) \b_{mj}^k - X_j (\a_m ) \b_{mi}^k - X_m (\a_k )
\b_{ij}^m \) \ = \ 0 \ . \end{array} $$
The first and last terms cancel each other due to $\b_{ji}^k = -
\b_{ij}^k$, and the term in square bracket vanishes due to the Jacobi
identity.
Using now \eqref{eq:Ssymcom} or \eqref{eq:sdeteq} to impose the
invariance of the DS \eqref{eq:gpds1}  under the first
$\sigma$-prolongation of the $X_i$ and using the fact that these are
standard symmetries of \eqref{eq:gpds2}, we obtain just the expression
\eqref{eq:gpsigma} for the $\sigma$, and then all hypotheses of Theorem 1 are
satisfied. \EOP

\bigskip

We stress that for this class of DS we have shown not only that the
determining equations for $\s$-symmetries can be solved, but have actually
provided a general class of solutions.

As a special case of \eqref{eq:gpds1} consider $f=Ax$ for some
matrix $A$; then obviously $\vphi_i=B_ix$, with $B_i$ matrices such that
$[A,B_i]=0$, provide standard symmetry vector fields $X_i$ for
\eqref{eq:gpds2}.
These matrices will satisfy $[B_i,B_j]=c^k_{ij}B_k$ and hence
$[X_i,X_j]=-c^k_{ij}X_k$, i.e. $\beta^k_{ij}=-c^k_{ij}$, and the vector
fields $X_i$ provide a $\s$-symmetry for the full DS
\[\dot x^a\ =\ (Ax)^a\ +\ \sum_{k=0}^s \a_k(x)(B_k x)^a\]
with $B_0=A$,  and with
\[\s_{ij}=\a_k(x)c^k_{ij}+(B_ix)^a(\pa \a_j(x)/\pa x^a)\ .\]

\section{Examples}

Our examples will be of small dimension $(n \le 4)$; thus we will slightly
change our notation about dependent coordinates, avoiding indices. The
original variables will be denoted by Latin
letters, in particular as $(x,y,z,w)$; the symmetry adapted ones by Greek
letters, in particular as
$(\xi , \eta , \chi )$ for variables entering in the reduced system, and
as $(\mu , \nu , \rho)$ for variables determined by the reconstruction
equations. To simplify notations we will often write $\pa_x$ instead of
$\pa/\pa x$, etc.

It can be noted that a comparison with what could be achieved in the
examples below by the use of {\it standard} symmetries is not easy nor
immediate, because one is not able in general to determine the full set
of standard symmetries of first order systems.

\subsection{Example 1.}

We start by considering a nearly-trivial case, i.e. with only one
$X$; in other words, this is a $\la$-symmetry (which is a special
case of $\s$-symmetry). We consider it in order to have a
specially simple case at hand, to look at through our present
approach. We consider a system which is {\it already} in
symmetry-adapted form (and coordinates), so that what matters here
is just the interpretation.

Let us consider the DS
\begin{eqnarray}
\xd &=& f(x,y) \nonumber \\
\yd &=& g(x,y) \label{eq:ex1a} \\
\zd &=& h(x,y) \ + \ f(x,y) \, z \nonumber \end{eqnarray}
with $f,g,h$ arbitrary smooth functions.

This admits as $\s$-symmetry the vector field $X = \pa_z$, with
$\s = \xd$; this implies that
\[ Y \ = \ \pa/\pa z \ + \ \xd \, \pa/\pa_\zd \ . \]
At order zero we have two invariants (apart from
the trivial one, $\tau = t$) for $X$ and hence for $Y$, i.e.
\[ \z_1 \ = \ x \ ; \ \ \z_2 \ = \ y \ . \]
The total derivatives of
these provide differential invariants of order one for $Y$, i.e.
\[ \z_1^{(1)} \ = \ D_t \, \z_1 \ = \ \xd \ , \ \ \z_2^{(1)} \ = \
D_t \, \z_2 \ = \ \yd \ . \]
There is a third differential invariant of order one,
\[ \b \ = \ \zd \ - \ \xd \ z \ . \]

Thus we have a set of (invariant) reduced equations, which are
just the first two of \eqref{eq:ex1a}, and a reconstruction
equation which is just the third one of these. Note that this
should be thought as defined on the solution to the first set; in
this sense the reconstruction equation can be rewritten as
\[ \b \ = \ k \ , \]
where $k$ should be seen as a constant on the
solutions to the reduced set (hence as an arbitrary function of
the invariants $\z_i, \z_i^{(1)}$). Indeed if we write this
equation in explicit form we have
\[ \zd \ - \ \xd \ z \ \ = \ k (x,y,\xd,\yd) \ ; \]
on the solutions to the reduced set this reads precisely (recall
the arbitrariness of $k$)
\[ \zd \ = \ f(x,y) \, z \ + \ k [x,y,f(x,y),g(x,y)] \ = \
h(x,y) \ + \ f(x,y) \, z \ . \]
Note once (and if) a solution $x(t)$, $y(t)$ to the reduced
equation is given, this reconstruction equation is a (non
autonomous) linear equation for $z(t)$,
\[ \zd \ = \ H(t) \ + \ F(t) \ z \ , \]
and hence can be solved.

According to Theorem 3, there is  $1=n-r-1$ constant of motion for the
DS \eqref{eq:ex1a} which is also invariant under $X=\pd/\pd z$. In this
case, the result
is trivial: it is enough to consider a constant of motion which depends
only on $x,y$.

\subsection{Example 2.}

We now consider a situation rather similar to the previous one, but with
a system which is not already in symmetry-adapted form.

Consider the three-dimensional DS
\begin{eqnarray}
\xd &=& - 2 \, z \ - \ x \, y^2 \, (x^2 + z^2)  \ + \ x \, y \, z \, \log(
y^2) \ , \nonumber \\
\yd &=& y^2 \ \( y \, (x^2 + z^2) \, - \, z \, \log (y^2) \) \ ,
\label{eq:exa2new3} \\
\zd &=& 2 \, x \ - \ y^2 \, z \,  (x^2 +  z^2) \ + \ y \, z^2 \, \log
(y^2) \ . \nonumber
\end{eqnarray}
If the system had a constant {\it in lieu} of the $\log (y^2)$ terms (or
if the argument was instead e.g. $x y$), it would admit as standard
symmetry the vector field
\[ X \ = \ x \, \pa_x \ - \ y \, \pa_y \ + \ z \, \pa_z \ ; \]
one can look for a $\s$ such that it is however a $\s$-symmetry. This is
the case, i.e. $X$ is a $\s$-symmetry (actually, having only one vector
field, this is a $\la$-symmetry), e.g. with
\[ \s \ = \ \xd \, y \ + \ x \, \yd \ = \ D_t \, (x \, y) \ . \]
In fact, the $\s$-prolonged vector field is in this case
$$ Y \ = \ X \ + \ \( \xd + x \xd y + x^2 \yd \) \, \frac{\pa}{\pa \xd} \
- \ \(  \yd + \xd y^2  + x y \yd \) \, \frac{\pa}{\pa \yd} \ + \ \( \zd +
\xd y z + x \yd z  \) \, \frac{\pa}{\pa \zd} \ , $$
and one can easily check that the system \eqref{eq:exa2new3} does indeed
admit $X$ as $\s$-symmetry.

Nontrivial invariants $\z_1,\z_2$ of order zero
(for $X$ and hence for $Y$ as well) are
\[ \z_1 \ = \ x \, y \ , \ \ \z_2 \ = \ y \, z  \]
which will be chosen as new (symmetry adapted) variables
$\xi=\z_1,\,\eta=\z_2$;
we can choose as additional variable $\rho = 1 + y^2$ in order to have
symmetry adapted coordinates. In the new coordinates,
\[ Y \ = \ 2 \, (1 - \rho) \ \frac{\pa}{\pa \rho} \ - \ 2 \, \( \dot{\rho}
- (1 - \rho) \dot{\xi} \) \ \frac{\pa}{\pa \dot{\rho}} \ . \]
It is immediate to check that $ \dot{\xi} = D_t \xi $ and $\dot{\eta} =
D_t \eta$ are differential invariants of order one. Moreover, we have an
additional differential invariant of order one; for this we can pick
\[ \mu \ = \ 2 \, (\yd / y) \ - \ (\xd y + x \yd) \ \log (y^2) \ . \]
Passing to the symmetry adapted variables, the system \eqref{eq:exa2new3}
is rewritten as
\[ \dot{\xi} =  - \, 2 \, \eta \ \ ,\
\dot{\eta}  =  2 \, \xi \ \ ;\
\dot{\rho}  =  2 \, (\xi^2 + \eta^2 )  \]
The first two represent the $\s$-symmetry reduced system, whose solution
is of course
\[ \xi (t) \ = \ \a \ \cos ( 2 t + \b ) \ , \ \
\eta (t) \ = \ \a \ \sin ( 2 t + \b )\ ; \]
the third one is the reconstruction equation, which on solutions to the
reduced system reads
\[ \dot \rho \ = \ 2 \, \a^2 \ ; \ \ \ \rho (t) \ = \ \rho_0  + 2 \,
\a^2 \ t \ . \]

A related example is obtained considering the three-dimensional DS
\begin{eqnarray*}
\xd &=& - x \ \( 1 \ + \ y \, (x - z) \ - \ x y^2 z \ \log (x z) \) \ ,
\nonumber \\
\yd &=& y \ \( 1 \ + \ y \, (x + z) \ - \ x y^2 z \ \log (x z) \) \ ,
 \\
\zd &=& z \ \( 1 \ - \ y \, (x + z) \ + \ x y^2 z \ \log (x z) \) \ .
\nonumber
\end{eqnarray*}
This system admits the same $\s$-symmetry (same $X$ and same $\sigma$) as
above,
and then the same  invariants and adapted coordinates,
in these it is rewritten as
\[\dot{\xi}  =  2 \, \xi \, \eta \ \ \dot{\eta}  =  2 \, \eta \ \
\dot{\rho}  =  - 2 \, \xi \]
The first two represent the $\s$-symmetry reduced system (which in this
case happens to be further reducible), whose solution is
\[ \xi (t) \ = \ \exp[(e^{2 t} - 1) \, \eta_0 ] \, \xi_0 \ , \ \
\eta (t) \ = \ e^{2 t} \, \eta_0 \ ; \]
the third one is the reconstruction equation, which on solutions to the
reduced system gives
\[ \rho (t) \ = \ \rho_0  - 2 \, \xi_0 \ \int_{t_0}^t \exp[(e^{2 \tau} -
1) \, \eta_0 ] \ d \tau \ . \]

According to Theorem 3, there is here  just one
constant of motion which is $X$-invariant: it is given by $I=x/z$.

\subsection{Example 3.}

This is an example of orbital reduction. The DS
$$ \dot x \, = \, yz \ , \
\dot y \, = \, z+xy \ , \
\dot z \, = \, w+xz \ , \
\dot w \, = \, y+xw $$
admits the two vector fields
$$ X_1\ = \ \frac{\pa}{\pa x} \ , \ \ X_2 \ = \ y \frac{\pa}{\pa y} \, + \, z \frac{\pa}{\pa z} \, + \, w \frac{\pa}{\pa w} $$
as a $\sigma$-orbital symmetry. Indeed this DS satisfies condition
\eqref{eq:Sorb} in the form
$$ [X_1,X_0]\ = \ z \, X_2 \ , \ \ [X_2,X_0] \ = \ X_0 + y z \, X_1 \ . $$
The two common invariants under $X_1,X_2$
$$ \zeta_1 \ = \ y/z \ , \ \ \zeta_2 \ = \ w/z $$
satisfy the orbitally reducible system
$$ \dot{\zeta}_1 \ = \ y \, \( \frac{1}{\zeta_1} \, - \, \zeta_2 \) \ , \ \
\dot{\zeta}_2 \ = \ y \, \(1 \, - \, \frac{\zeta_2^2}{\zeta_1} \) \ , $$
in agreement with Remark 3. The reduced equation is
$$ \frac{\d \zeta_1}{\d \zeta_2} \ = \ \frac{1 - \zeta_1 \zeta_2}{\zeta_1 - \zeta_2^2}
\ . $$

\subsection{Example 4.}

As stated in Section \ref{sec:det_gen}, finding the most general DS which
admits a given $\s$-symmetry is often feasible. We consider the two
(commuting) vector fields,
\begin{eqnarray*} X_1
&=& - 2 \, z \, \frac{\pa}{\pa x} \ + \ \frac{\pa}{\pa
z} \ , \\ X_2 &=& 8 y z w  \, \frac{\pa}{\pa x}
\ + \ 2 w \, \frac{\pa}{\pa y} \ - \ 4 y w \,
\frac{\pa}{\pa z} \ + \ \frac{\pa}{\pa w} \ ; \end{eqnarray*}
and choose
\[ \s \ = \ \pmatrix{0 & \dot{x} + 2 z \dot{z}  \cr
\dot{y} - 2 w \dot{w} & 0 \cr} \ . \]
The $\s$-prolonged vector fields will be written as $(k=1,2)$
\[ Y_k \ = \  \phi_k^i \, (\pa / \pa x_i) \ + \
\psi_k^i (\pa / \pa \dot{x}_i) \ , \]
where the prolongation coefficient vectors are
\begin{eqnarray*}
\psi_1 &=& (-2 \zd + 8 A y z w ,
  2 A w , -4 A y w , A) \ , \\
\psi_2 &=& ( 8 y w \zd + 2 B z , 2 \wdot , -
  B , 0) \ ; \end{eqnarray*}
here $A = 2 z \zd + \xd$, $B = (4 w - 1) \yd +
2 (2 y + w ) \wdot $. Note that $[X_1,X_2] = 0$ and $[Y_1,Y_2]=0$.

There are two nontrivial common invariants of order zero, i.e.
\[ \z_1 \ = \ x \ + \ z^2 \ , \ \ \z_2 \ = \ y - w^2 \ ; \]
and  correspondingly we get first order differential invariants
\[ \z_1^{(1)} \ = \ \xd \ + \ 2 \, z \, \zd \ , \ \
\z_2^{(1)} \ = \ \yd \ - \ 2 \, w \, \wdot \ . \]
The two additional first order invariants are
\[ \b_1 \ = \ \wdot \ - \ (z + y^2) \, (\xd + 2 z \zd ) \ , \ \
\b_2 \ = \ \zd \ + \ 2 \, y \, \yd \
- \ w (\yd - 2 w \wdot ) \ . \]
We take as symmetry adapted coordinates
\[ \xi  = \z_1  =   x + z^2 \ , \ \ \eta = \z_2 = y - w^2 \ , \ \ \mu = z +
y^2 \ , \ \ \nu = w \ ; \]
note the first two are
invariants, while the other two rectify the $X_i$. With
these coordinates we have of course
$\z_1^{(1)} =
\dot{\xi}$, $\z_2^{(1)} = \dot{\eta}$; moreover
\[ \b_1 \ = \ \dot{\nu} \ - \ \dot{\xi} \, \mu \ , \ \ \b_2 \ = \
\dot{\mu} \ - \ \dot{\eta} \, \nu \ . \]
The vector fields read now $X_1 = \pa_\mu$, $X_2 = \pa_\nu$; as in these
coordinates
\[ \s \ = \ \pmatrix{0 & \dot{\xi} \cr \dot{\eta} & 0 \cr} \ , \]
the $\s$-prolonged ones are just
\[ Y_1 \ = \ (\pa / \pa \mu ) \ + \ \dot{\xi} \, (\pa / \pa \dot{\nu} ) \
, \ \
Y_2 \ = \ (\pa / \pa \nu ) \ + \ \dot{\eta} \, (\pa / \pa \dot{\mu} ) \ .
\]
Thus any system of the form
\begin{eqnarray}
\xd &=& [1 + 8 y z w (y^2 + z) ] F + 2 z [(2 y - w) G - H + 4 y w K]
\nonumber \\
\yd &=& 2 (y^2 + z) w F + G + 2 w K \nonumber \\
\zd &=& - 4 y w (y^2 + z) F + (w - 2 y) G + H - 4 y w K \label{eq:exa3g} \\
\wdot &=& (y^2 + z) F + K \nonumber \end{eqnarray}
where we have noted $F = f(\z_1 , \z_2)$, $G = g(\z_1 , \z_2)$, $H =
h(\z_1 , \z_2)$, $K = k(\z_1 , \z_2)$, admits $\{ X_1 , X_2 \}$ as
$\s$-symmetry with the $\s$ considered above (this is not the most
general case, but is the most general -- up to multiplication by nowhere
vanishing functions \cite{CDW} -- one if we require to have polynomial
functions; in this case $f,g,h$ should of course be polynomial).

As stated by our Theorems, any system in the class \eqref{eq:exa3g} can be
reduced via $\s$-symmetry reduction. In fact, passing to symmetry adapted
coordinates as above yields the system \eqref{eq:exa3g} in the form
\begin{eqnarray*}
\dot{\xi} &=& f(\xi , \eta ) \ , \\
\dot{\eta} &=& g (\xi , \eta ) \ ; \\
\dot{\mu} &=& h (\xi , \eta ) \ + \ g(\xi,\eta) \ \nu \ , \\
\dot{\nu} &=& k (\xi , \eta ) \ + \ f(\xi,\eta) \ \mu \ . \end{eqnarray*}
The first two equations represent the reduced system, while the last two
are the reconstruction equations.

The latter can be written, on solutions to the reduced system, in the form
$\b_i = \kappa_i (\z_1,\z_2,\z_1^{(1)},\z_2^{(1)} ) $ ($i=1,2$) for a
suitable choice of the functions $\kappa_i$. In fact, using the explicit
form of $\b_i$ we would have
\[ \dot{\nu} \ - \ \dot{\xi} \, \mu \ = \ \kappa_1
(\xi,\eta,\dot{\xi},\dot{\eta}) \ , \ \
\dot{\mu} \ - \ \dot{\eta} \, \nu \ = \ \kappa_2
(\xi,\eta,\dot{\xi},\dot{\eta}) \ ; \]
on the solutions to the reduced equations these read
\begin{eqnarray*}
\dot{\nu} &=& \kappa_1 [\xi,\eta,f(\xi,\eta),g(\xi,\eta) ] \ + \
f(\xi , \eta) \, \mu \ = \ k (\xi,\eta) \ + \ f(\xi,\eta) \, \mu \ , \\
\dot{\mu} &=& \kappa_2 [\xi,\eta,f(\xi,\eta),g(\xi,\eta) ] \ + \
g(\xi,\eta) \, \nu \ = \
h(\xi,\eta) \ + \ g(\xi,\eta) \, \nu \ . \end{eqnarray*} On a given
solution
$\{\xi(t),\eta(t)\}$ to the reduced equations, these are a system of
non autonomous linear equations
\[ \dot{\mu} \ = \ H(t) \ + \ G(t) \ \nu \ , \ \ \ \dot{\nu} \ = \ K(t) \
+ \ F(t) \ \mu \ , \]
in agreement to our general discussion.

As a concrete example, one can consider
\begin{eqnarray*}
\xd &=& 1 \ + \ 4 \, y \, z \ - \ 2 \, w \, z \ (1 - 4 y z - 4 y^3) \\
\yd &=& 1 \ + \ 2 \, w \ (z + y^2)  \\
\zd &=& w \ - \ 2 \, y \ - \ 4 \, y \, w \ (z + y^2 )  \\
\wdot &=& z \ + \ y^2 \ .   \end{eqnarray*}
In the adapted coordinates, this reads
\[
\dot{\xi}  = 1 \ , \ \dot{\eta} = 1 \ ; \
\dot{\mu}  = \nu  \ , \ \dot{\nu}  = \mu  \ . \]

In agreement with Theorem 3, there is one $X$-invariant constant of
motion $(n=4,r=2)$, which is given by $I=x-y+z^2+w^2$.

\subsection{Example 5.}

Let us now consider vector fields
\[ X_1 \ = \ \frac{1}{1 + 2 x} \ (\pa_y \ - \ \pa_x ) \ , \ \ X_2 \ = \
\pa_z \ ; \]
and the matrix \[ \s \ = \ \pmatrix{ 0 & \xd + \yd \cr \xd +
\yd & 0 \cr } \ . \]

The $\s$-prolonged vector fields are \begin{eqnarray*} Y_1 &=& X_1 \ +
\ \frac{1}{(1 + 2 x)^2} \ \left[ 2 \xd \, \( \frac{\pa}{\pa
\xd} \, - \, \frac{\pa}{\pa \yd} \) \, + \, (1 + 2
x)^2 \, (\xd + \yd ) \, \frac{\pa}{\pa \zd} \right] \ , \\
Y_2 &=& X_2 \ + \ \frac{(\xd - \yd)}{1 + 2 x} \(
\frac{\pa}{\pa \yd} \ - \ \frac{\pa}{\pa \xd} \) \ .
\end{eqnarray*}
It is immediate to check that $[X_1,X_2] = 0$, $[Y_1,Y_2]=0$.

The most general system admitting $\X = \{X_1 , X_2 \}$ as $\s$-symmetry
is easily determined, and is given by
\begin{eqnarray}
\xd &=& - \, (1+2 x)^{-1} \ [ (1 - x - y - z) \, F \ + \ G
] \ , \nonumber \\
\yd &=& (1+2 x)^{-1} \ [ (3 x + y + z) \, F \ + \ G ] \ ,
\label{eq:ex4a} \\
\zd &=& - \, (1 - y + x^2) \, F \ + \ H \ , \nonumber
\end{eqnarray}
where $ F =  f (x+y )$, $ G = g ( x+y)$, $H = h (x+y )$.

The nontrivial invariant of order zero for $X_1,X_2$, and hence also for
$Y_1 , Y_2$, is
\[ \z \ = \ x \ + \ y \ ; \]
it follows from our general discussion that the total derivative
of this, i.e. $\z^{(1)} = \dot{\xi} + \dot{\eta}$, is a first order
differential invariant, and indeed one checks easily this is the case.
There are two
additional first order invariants, which are
\[ \b_1 \ = \ \yd \ - \ z \, \xd \ , \ \ \b_2 \ = \ \zd \ - \ y \,
\xd \ . \]

Passing to symmetry-adapted coordinates
\[ \xi \ =\z_1 \ =  \ x + y \ , \ \ \mu \ = \ y - x^2 \ , \ \ \nu \ = \
x + y + z \ , \]
the basic and $\s$-prolonged vector fields read
\begin{eqnarray*} X_1 &=& \pa_\mu \ , \ \ X_2 \ = \ \pa_\nu \ ; \\
Y_1 &=& (\pa/ \pa \mu) \, + \, \dot{\xi} (\pa / \pa \dot{\nu}) \
, \ \ Y_2 \ = \ (\pa / \pa \nu) \, + \, \dot{\xi} \, (\pa / \pa \dot{\mu} )
\ . \end{eqnarray*}
The general system \eqref{eq:ex4a} reads, in these coordinates,
\begin{eqnarray*}
\dot{\xi} &=& f(\xi ) \ ,   \\
\dot{\mu} &=& f(\xi) \, \nu \ + \ g(\xi) \ ,  \\
\dot{\nu} &=& f(\xi) \, \mu \ + \ h(\xi) \ ;
\end{eqnarray*}
The reduced system is just the first equation, the
other two representing the reduction equations; these are indeed
of the form $\b_1 = g(\z)$, $\b_2 = h(\z)$.

Note that in this (and the following) example, Theorem 3 does not admit
the presence of $\s$-invariant constants of motion: we have indeed
$n-r-1=0$.

A more concrete case is obtained e.g. by choosing
\[ f(x) = x - x^3 \ , \ \ g(x) \ = \ x \ , \ \ h(x) = x^2 \ . \]
With these choices, the original system has the quite involved expression
\begin{eqnarray*}
\xd &=& - (1+ 2 x)^{-1} \, [ (x + y)^3 \, (1 - x -
y - z) \ + \ (x+y) \, (x+y+z) ]
\ , \nonumber \\
\yd &=&  (1+ 2 x)^{-1} \, [  (x+y) \(1 -(3 x +
y ) (x + y)^2 \right. +   \\
& &\left.  \ + \ 3 x + y
+ z  (1 - x - y) (1 + x + y) \) ]\ , \\
\zd &=& (x+y) \, (y - 1 - x^2 ) \ + \ (x + y)^2 \ + \
(1 + x)^3 \, (1 - y + x^2) \ .
\end{eqnarray*}
Passing to symmetry adapted coordinates this system  reads
\begin{eqnarray*}
\dot{\xi} &=& \xi \ - \ \xi^3 \ ,  \\
\dot{\mu} &=& (\xi - \xi^3) \, \nu \ + \ \xi \ ,  \\
\dot{\nu} &=& (\xi - \xi^3) \, \mu \ + \ \xi^2 \ .
\end{eqnarray*}

\subsection{Example 6.}

Let us now consider a situation with a reduced system of
dimension one. We consider the four-dimensional DS
\begin{eqnarray}
\xd &=& - (x + z^2) + e^{- (y + z)} \! +\! 2 z \( e^{- y + w^2} + 2 e^{-
w}
w - 2 e^{- (y+z)} \)  (x + z^2)  \ , \nonumber \\
\yd &=& \( e^{- y + w^2} + 2 e^{- w} w \) \, (x + z^2 ) \ , \nonumber \\
\zd &=& - \(e^{- y + w^2} + 2 e^{- w} w - 2 e^{- (y + z)} \) (x + z^2) \ -
\ e^{- (y + z)} \ , \label{eq:ex5} \\
\wdot &=& e^{- w} \, (x + z^2 ) \ . \nonumber \end{eqnarray}

We are apparently clueless in front of such an involved DS, and this is
maybe the appropriate place to show how one can proceed in trying to
determine $\s$-symmetries for such a system. The recurring term $(x +
z^2)$ suggests to look for vector fields which admit this term as an
invariant. This requirement just selects vector fields $X = \phi^a \pa_a$
such that $\phi^1 = - 2 z \phi^3$; we obviously have three functionally
independent such fields, and we can e.g. choose the vector fields
\begin{eqnarray*}
X_1 &=& 2 z \pa_x + \pa_y - \pa_z \ ,  \ \ X_2 \ = \ - 2 z \pa_x + \pa_z \
, \\
X_3 &=& 4 z w \pa_x + 2 w \pa_y - 2 w \pa_z + \pa_w \ ; \end{eqnarray*}
note that these $X_i$ have been chosen in order to be autonomous, simple,
and to commute with each other; moreover, the rectifying change of
coordinates will have a Jacobian with unit determinant.

We can then look for a $\s$ matrix satisfying the $\s$-determining
equations \eqref{eq:sdeteq}; we have thus a system of twelve equations for
the
nine functions $\overline{\s}_{ij}$. In this case one obtains by simple
linear algebra that $\overline{\s}$ must be diagonal, and more precisely
\[ \overline{\s} \ = \ \mathrm{diag} \{ - e^{y - w^2} (x + z^2) \, , \,
e^{- (y+z)} (1 - 2 (x+z^2)) \, , \, - e^{-w} (x + z^2) \} \ . \]
This
expression is not very nice, but using the equations of motion
\eqref{eq:ex5} we can rewrite
$$e^{- y + w^2}  =  \( \frac{\yd - 2 w \wdot}{x+z^2} \)_S \ , \ \
e^{- (y+z)}  =  - \, \( \frac{\yd + \zd}{1 - 2 (x+z^2)} \)_S \ , \ \
e^{- w} =  \( \frac{\wdot}{x+z^2} \)_S $$
(where as usual $S=S(\E)$ denotes the
restriction to the solutions of the system); with these, we can write
\[ \s \ = \ - \ \mathrm{diag} \{ \yd - 2 w \wdot \, , \, \yd + \zd \, , \,
\wdot \} \ = \ D_t \, \left[ - \mathrm{diag} \{ y - w^2 , y + z , w \}
\right] \ .
\]
One can quite easily evaluate the $\s$-prolonged vector fields $Y_i$
and verify that also commute with each other.
The only common invariant of order zero for the $X_i$ (and hence also for
the $Y_i$) is
\[ \z \ = \ \xi \ = \ x \, + \, z^2 \ . \]
We can complete the system of coordinates by defining
\[ \mu \ = \ y \ - \ w^2 \ , \ \ \nu \ = \ y \ + \ z \ , \ \ \rho \ = \ w
\ ; \]
this yields a Jacobian with unit determinant.

The total derivative of $\z$ provides a differential invariant
\[ \z^{(1)} = D_t \z \ = \ \xd \ + \ 2 z \zd \ . \]
There are three additional first order differential invariants, which can
be chosen as
\begin{eqnarray*}
\b_1 &=& (\yd \ - \ 2 \, w \wdot) \ \exp (y - w^2) \ , \\
\b_2 &=& (\yd + \zd) \ \exp (y + z) \ , \\
\b_3 &=& \wdot \ \exp (w) \ . \end{eqnarray*}

In the symmetry-adapted coordinates, the basic vector fields read
\[ X_1 = \pa_\mu \ , \ \ X_2 = \pa_\nu \ , \ \ X_3 = \pa_\rho \ , \]
so that these are indeed rectifying coordinates; in these same
coordinates,
the $\s$-prolonged vector fields read
\[ Y_1 = \pa_\mu \ - \ \dot{\mu} \, \pa_{\dot{\mu}} \ , \ \
Y_2 = \pa_\nu - \dot{\nu} \pa_{\dot{\nu}} \ , \ \ Y_3 = \pa_\rho \ - \
\dot{\rho} \, \pa_{\dot{\rho}} \ . \]
Note that the $\b_i$ do now read simply
\[ \b_1 = \dot{\mu} e^{\mu} \ , \ \ \b_2 = \dot{\nu} e^{\nu} \ , \ \ \b_3
= \dot{\rho} e^{\rho} \ . \]

Finally, the DS under study reads in these variables
\begin{eqnarray*}
\dot{\xi} &=& - \xi \ ,  \\
\dot{\mu} &=& e^{- \mu} \ \xi \ ,  \\
\dot{\nu} &=& e^{- \nu} \ (2 \xi - 1)  \ ,  \\
\dot{\rho} &=& e^{- \rho} \ \xi \ .   \end{eqnarray*}
We thus have a one-dimensional reduced system (the equation for
$\dot{\xi}$) and three reconstruction equations; these can be written as
$ \b_i = h_i (\xi )$, as our discussion shows to be always the case; in
this case we have $h_1 (\xi ) = h_3 (\xi) = \xi$, $h_2 (\xi) = 2 \xi - 1$.

\subsection{Example 7.}

In the examples discussed so far, we were in the situation
considered by Theorem 1, i.e. $\X$ and $\Y$ had the same
involution relations, and hence in particular the same rank. We
will now consider a case within the framework of Theorem 2.

Let us consider the (commuting) vector fields
\[ X_1 \ = \ \pa_y \ , \ \ X_2 \ = \ 2 z \, \pa_x \ - \ (x + 2 z^2) \,
\pa_y \ + \ \pa_z \ ; \]
and the matrix
\[ \s \ = \ \pmatrix{ \xd - 2 z \zd & 0 \cr 0 & \yd + x \zd + z \xd \cr} \
. \]
The $\s$-prolonged vector fields are
\begin{eqnarray*}
Y_1 &=& X_1 \ + \ (\xd - 2 z \zd) \, (\pa / \pa \yd ) \ ; \\
Y_2 &=& X_2 \ + \ 2 \(\zd + z (\yd + x \zd + z \xd)\) \, (\pa / \pa \xd )
\\
& & \ - \ \(\xd + 4 z \zd + (x + 2 z^2) (\yd + x \zd + z \xd) \)  \, (\pa
/
\pa \yd ) \\
& & \ + \
 (\yd + x \zd + z \xd)    \, (\pa / \pa \zd ) \ . \end{eqnarray*}
These do not commute; we have instead
$$ [Y_1,Y_2] \ = \ Y_3 \ = \
2 z (\xd - 2 z \zd)  \, (\pa / \pa \xd ) \ + \ (x + 2 z^2) (2 z \zd -
\xd)  \, (\pa / \pa \yd ) \ + \ (\xd - 2 z \zd)  \, (\pa / \pa \zd ) \ . $$
It is easy to check that (as stated in our general
discussion, see Section \ref{sec:completion}) the auxiliary vector
field $Y_3$ commutes with the other ones: $[Y_1,Y_3] = 0 = [Y_2,Y_3]$.

With the notation of Section \ref{sec:completion}, we have $n=3$,
$n_0=2$, $r_0 = 2$, $r=3$; it follows that the reduced DS has dimension
$\kappa_0 = n - r_0 = 1$, and the number of reconstruction equations is
$\theta = r_0 - \delta = 1$.

The nontrivial invariant of order zero and the corresponding first order
differential invariant are
\[ \z \ = \ \xi \ = \ x - z^2 \ ; \ \ \z^{(1)} \ = \ D_t \z \ = \ \xd \ -
\ 2 \ z \ \zd \ ; \]
there is an additional first order invariant, which turns out to be
\[ \b \ = \ (\yd  + x \zd + z \xd) - (y + x z) (\xd - 2 z \zd) \ . \]

Any system of the form
\begin{eqnarray}
\xd &=& f(x - z^2) \ + \ 2 z \zd \ , \nonumber \\
\yd &=& - z f(x - z^2) + (y + x z) f(x - z^2) + g(x - z^2) - x \zd
\label{eq:exa6} \end{eqnarray}
is (by construction) invariant under the system $\overline{\Y} = \{ Y_1 ,
Y_2 , Y_3 \}$ of vector fields; this is the completion of $\Y$, the
$\s$-prolongation of $\X = \{ X_1 , X_2 \}$. Note that in \eqref{eq:exa6}
the variable $z$ should be considered as a parameter; it can change in
time, with speed $\zd$; or we can set this to zero, obtaining a slightly
less general set of systems
\begin{eqnarray*}
\xd &=& f(x - z^2)  \ , \\
\yd &=& - z f(x - z^2) + (y + x z) f(x - z^2) + g(x - z^2) \ .
\end{eqnarray*}

The adapted coordinates for the vector fields we are considering are
\[ \xi = x - z^2 \ , \ \ \eta = y + x z \ ; \ \ \mu = z \ . \]
(Note $\b = \dot \eta - \eta \dot \z$.)
In terms of these we have $X_1 = \pa_\eta$, $X_2 = \pa_\mu$, and
\begin{eqnarray*}
Y_1 &=& (\pa / \pa \eta) \ + \ \dot{\xi} \, (\pa / \pa \dot{\eta} ) \ , \\
Y_2 &=& (\pa / \pa \mu) \ + \ \dot{\eta} \, (\pa / \pa \dot{\mu} ) \ , \\
Y_3 &=& \dot{\xi} \, (\pa / \pa \dot{\mu} ) \ . \end{eqnarray*}
In these variables, the system \eqref{eq:exa6} reads
\begin{eqnarray*}
\dot{\xi} \ = \ f(\xi ) \ ,  \\
\dot{\eta} \ = \ g ( \xi ) \ + \ \eta \ f(\xi ) \ .
\end{eqnarray*}
Needless to say, the first equation
represents the reduced system, while the second
is the reconstruction equation; on solutions to the former one,
 the latter can also be written as
$ \beta = g (\xi )$.

\subsection{Example 8.}

Finally, let us consider an example of the situation dealt with in Section
\ref{sec:det_gp} (see in particular Theorem 4), i.e. the family of DS
\begin{eqnarray}
\xd &=& x + a y + \a_1(x,y,z) x + \a_2(x,y,z) y \nonumber \\
\yd &=& a x + y + \a_2(x,y,z) x + \a_1(x,y,z) y \label{eq:exa7} \\
\zd &=& b z + \a_1(x,y,z) z \nonumber \end{eqnarray}
with $a,b$ constants and $\a_1, \a_2$ smooth functions. The
(commuting) vector fields
\[ X_1 \ = \ x \, \pa_x \ + \ y \, \pa_y \ + \ z \, \pa_z \ , \ \
X_2 \ = \ y \, \pa_x \ + \ x \, \pa_y   \]
are standard symmetries for the linear part of the above DS; note there
would also be another standard symmetry for this linear DS, as $z \pa_z$
and $x \pa_x + y \pa_y$ are separately symmetries for it, but only $X_1$
and $X_2$ enter in the nonlinear part of the DS in the way discussed in
Section \ref{sec:det_gen}, with $\a_i$ associated to $X_i$.

The vector fields $X_1,X_2$ are standard symmetries of the full DS
\eqref{eq:exa7} only if the $\a_i (x,y,z)$ depend uniquely on their common
invariant $\z = (x^2-y^2)/z^2$; we consider the case where they are
instead arbitrary smooth functions.

According to Theorem 4, they are a $\s$-symmetry for the full system
with $\s$ given by equation \eqref{eq:gpsigma}; as $[X_1,X_2] = 0$,
this equation  yields simply $ \s_{ij} \ = \ X_i (\a_j )$
or
\[ \s \ = \ \pmatrix{(\a_1)_x x + (\a_1)_y y + (\a_1)_z z &  (\a_2)_x x +
(\a_2)_y y + (\a_2)_z \ z \cr  (\a_1)_y x + (\a_1)_x y & (\a_2)_y x +
(\a_2)_x y \cr} \ . \]
With this choice of $\s$, some standard algebra provides the
$\s$-prolonged vector fields $Y_1,Y_2$ and shows
that these are indeed $\s$-symmetries for the
system \eqref{eq:exa7}, for any choice of the smooth functions $\a_1$ and
$\a_2$.

According to Theorem 1, one can introduce the common invariant
\[ \zeta \ = \xi= \  (x^2 \, - \, y^2) \, / \, z^2 \]
and then obtain the reduced equation, which should be in the form
$ \dot\xi  = f ( \xi )$; indeed, using \eqref{eq:exa7} we obtain
easily that
\[ \dot\xi \ =  \ 2 \, (1-b) \ \xi \ . \]
The additional first order invariants and the reconstruction equations
would of course depend on the functions $\a_i$.

We specialize this example to a concrete DS considering
\begin{eqnarray}
\xd &=& x - y - x^2 y - y z^2  \nonumber \\
\yd &=& y - x - x (y^2 + z^2)   \label{eq:exa8}\\
\zd &=& 2 z - x y z \nonumber \nonumber \end{eqnarray}
In this case we get
\[ \s \ = \ \pmatrix{- 2 x y & - 2 z^2 \cr - (x^2 + y^2) & 0 \cr} \ . \]
The $\s$-prolonged vector fields are easily obtained and
again one can check  that these are indeed $\s$-symmetries for the
system \eqref{eq:exa8}. Beside the invariants $\z = (x^2 - y^2)/z^2$ and
$\z^{(1)} = D_t \z $, one has the additional first order differential
invariants
\[ \mu \ = \ \frac{(x^2 - y^2) z^2 + (x \yd - y \xd)}{x^2 - y^2} \ , \ \
\nu \ = \ \frac{x y (x^2 - y^2) + (x \xd - y \yd)}{x^2-y^2} \ . \]
The system \eqref{eq:exa8} is then rewritten as
\[
\dot \xi \ = \ - \, 2 \ \xi \ ;  \ \
\mu \ = \ - \, 1 \ , \ \
\nu \ = \ 1 \ . \]
The first of this represents the reduced DS, the latter two the
reconstruction equations.

\section{Conclusions}

In a recent work \cite{Sprol} we have introduced a modification of the prolongation operation on sets of vector fields, called $\s$-prolongation; and correspondingly the concept of $\s$-symmetries for systems of ODEs. The formulation of that paper was not able to deal with the special but relevant case of Dynamical Systems (systems of first order ODEs). The aim, and the main result, of this paper is to  extend our approach to encompass also the case of Dynamical Systems.

We were able to deal both with the case where the original and the $\s$-prolonged vector fields generate a foliation of the same rank (Theorem 1) and the case where the ranks differ (Theorem 2); in both cases one is led to study a reduced system together with a set of reconstruction equations. The latter, at difference with the case met in the reduction based on standard symmetries, do not amount to quadrature but are a set of first order ODEs, which might well be hard to solve in practice.

Our approach focuses on foliations generated by symmetry (in this case, $\s$-symmetry) vector fields; it is thus entirely natural that it falls within the general Frobenius theory. In this sense, a $\s$-symmetry for a dynamical vector field $X_0$ with (standard) prolongation $Y_0$ is a set of vector fields $X_i$ whose $\s$-prolongations $Y_i$ span a foliation $\mathcal{F}$ such that $Y_0$ commutes with vector fields in $\mathcal{F}$ modulo $\mathcal{F}$ itself. This situation can also be described by saying that $Y_0$ is projectable with respect to the foliation $\mathcal{F}$ \cite{MM}.

It is thus natural that a direct application of the classical Frobenius theorem provides the relation between $\s$-symmetries and constants of motion (Theorem 3).

The actual determination of $\s$-symmetries can be a highly nontrivial task, as the determining equations for these do not share the nice properties of those for standard symmetries. For dynamical systems with a special structure, one can try to use this to relate $\s$-symmetries to standard symmetries of a related (maybe simpler) system. We provided some constructive result in this direction (Theorem 4).

It should be mentioned that our theory is able to deal not only with reduction, but also with orbital reduction \cite{HaWa,Wal99}; again this is quite natural once one adopts the Frobenius point of view, as foliations and their integral manifolds (or curves) do not depend on the parametrization of vector fields and curves.

In conclusion, dealing with dynamical systems allowed to have a clearer geometrical picture with respect to the general case (considered in previous work \cite{Sprol}); this suggests that one should focus on foliations generated by symmetry -- or orbital symmetry -- vector fields rather than on the individual symmetry vector fields; at the algebraic level, this suggests to focus on the Lie  module structure rather than just on the Lie algebra one (see also \cite{CGW}).

As already mentioned in the Introduction (and in \cite{Sprol}), our approach should be seen as a development of the approach by Pucci and Saccomandi
\cite{PuS} to the Muriel and Romero beautiful and fruitful idea of $\la$-symmetries \cite{MuRom1}.

\medskip\noindent
{\bf Acknowledgements.} We thank  an unknown referee for pointing
out the relations to projectable vector fields. The research of GG is partially supported by MIUR-PRIN program under project 2010-JJ4KPA.


\end{document}